\documentclass[aps,twocolumn,showpacs,psfig,superscriptaddress,longbibliography]{revtex4-2} % APS journal format with two-column layout
\usepackage{amsfonts}        % AMS mathematical fonts
\usepackage{mathrsfs}        % Ralph Smith's Formal Script font for math
\usepackage{amsmath}         % needed for subequations
\usepackage{color}           % basic color support
\usepackage{natbib}          % natural bibliography citations
\usepackage{textcomp}        % text companion fonts and symbols
\usepackage{graphicx}        % enhanced graphics inclusion
\usepackage{bm}              % bold maths
\usepackage{amssymb}         % AMS mathematical symbols

\usepackage{xspace}          % intelligent spacing after macros
\usepackage{epstopdf}        % automatic EPS to PDF conversion
\usepackage{dcolumn}         % Align table columns on decimal point
\usepackage{longtable}       % tables that span multiple pages
\usepackage{multirow}        % merge table rows vertically
\usepackage[colorlinks=true, letterpaper=true, pdfstartview=FitV, linkcolor=blue, citecolor=blue, urlcolor=blue]{hyperref} % hyperlinks with blue colors
\usepackage{braket}          % Dirac bra-ket notation for quantum mechanics
\usepackage{float}           % improved float positioning with [H] option
\usepackage{siunitx}         % to insert the percentage sign

\usepackage{xcolor}
\definecolor{revcolor}{RGB}{0,0,0}
\newcommand{\rev}[1]{\textcolor{revcolor}{#1}}

\begin{document}

\title{Many-Body Correlation Effects in Fr\"ohlich Electron-Phonon Coupling}

\author{Zien Zhu}
\affiliation{Mork Family Department of Chemical Engineering and Materials Science, University of Southern 
California, Los Angeles, CA 90089, USA}

\author{Chih-En Hsu}
\affiliation{Mork Family Department of Chemical Engineering and Materials Science, University of Southern 
California, Los Angeles, CA 90089, USA}
\affiliation{Department of Physics, Tamkang University, Tamsui, New Taipei 251301, Taiwan}

\author{Benran Zhang}
\affiliation{Mork Family Department of Chemical Engineering and Materials Science, University of Southern 
California, Los Angeles, CA 90089, USA}

\author{Zhenfa Zheng}
\affiliation{Mork Family Department of Chemical Engineering and Materials Science, University of Southern 
California, Los Angeles, CA 90089, USA}

\author{Mauro Del Ben}
\affiliation{Applied Mathematics and Computational Research Division, Lawrence Berkeley National Laboratory, Berkeley, California 94720, USA}

\author{Antonios M. Alvertis}
\affiliation{Oden Institute for Computational Engineering and Sciences, The University of Texas at Austin, Austin, Texas 78712, USA}
\affiliation{Department of Physics, The University of Texas at Austin, Austin, Texas 78712, USA}

\author{\\Hung-Chung Hsueh}
\affiliation{Department of Physics, Tamkang University, Tamsui, New Taipei 251301, Taiwan}

\author{Zhenglu Li}
\email{zhenglul@usc.edu}
\affiliation{Mork Family Department of Chemical Engineering and Materials Science, University of Southern 
California, Los Angeles, CA 90089, USA}

\begin{abstract}
In compound semiconductors and insulators, the polar electron-phonon coupling diverges at long range, known as the Fr\"ohlich interaction. 
Modern first-principles electron-phonon calculations treat the Fr\"ohlich interaction in a semiclassical electrostatic formalism based on density-functional perturbation theory. 
Here, using many-body $GW$ perturbation theory, we reveal important electron correlation effects in the Fr\"ohlich-type electron-phonon coupling, which are missed by the prevailing approaches and rooted in the fundamentally distinct behavior between quasiparticle self-energy and semi-local exchange-correlation functionals in the long-range limit. 
Going beyond the electrostatic treatment, we derive and implement the $GW$ self-energy contribution to the long-range polar electron-phonon coupling, and demonstrate its critical role and nontrivial behaviors in properties such as electron linewidth and polaron formation in several prototype semiconductors. Remarkably, calculations on photoemission kink in electron-doped TiO$_2$ achieve excellent agreement with experiment. 
Our work establishes the many-body generalization of the Fr\"ohlich interaction that is essential for accurate electron-phonon calculations at the full $GW$ level combined with Wannier interpolation techniques.
\end{abstract}

\maketitle

The Fr\"ohlich interaction between electrons and longitudinal optical (LO) phonons diverges in the long-wavelength limit (phonon wavevector $\textbf{q} \rightarrow 0$) in compound semiconductors and insulators \cite{frohlich1954electrons}, where the Born effective charges are nonzero \cite{born1996dynamical}. 
The divergence is rooted in the insufficient screening (unlike in metals) of the Coulomb dipole fields from the atomic oscillations, and plays a critical role in quasiparticle lifetimes \cite{nery2018quasiparticles}, zero-point renormalization \cite{miglio2020predominance}, transport \cite{lee2020ab}, optical absorption \cite{marini2024optical}, photoluminescence \cite{wright2016electron}, and polaron formation \cite{adamowski1989formation, de2023high}, among others.
The standard first-principles approaches based on density-functional perturbation theory (DFPT) \cite{baroni2001phonons}  and Wannier interpolation techniques \cite{marzari2012maximally} compute electron-phonon ($e$-ph) coupling by separating it into short-range (SR) and long-range (LR) parts \cite{verdi2015frohlich, sjakste2015wannier}.
The SR $e$-ph coupling can naturally be accurately interpolated by Wannier functions, whereas the LR Fr\"ohlich interaction is often treated explicitly by evaluating the $e$-ph matrix elements of a semiclassical dipole-induced electrostatic potential \cite{verdi2015frohlich, sjakste2015wannier,deng2021ab, sohier2016two}, which can be further extended to quadrupolar interactions \cite{jhalani2020piezoelectric, park2020long, brunin2020electron, brunin2020phonon}. These established methods capture the asymptotic behavior  $\sim 1/|\textbf{q}|$ in the dipolar Fr\"ohlich $e$-ph coupling quite effectively, providing reliable first-principles calculations of $e$-ph coupling in semiconductors and insulators at the density functional theory (DFT) level \cite{zhang2022phonon,ma2018first,ranalli2024electron,ponce2018towards,ponce2021first,ponce2023accurate,ponce2023long}.

The recently developed $GW$ perturbation theory ($GW$PT) method \cite{li2019electron,li2024electron} offers a many-body-level description of the $e$-ph coupling, going beyond DFPT. 
It has been demonstrated (via frozen-phonon $GW$ and $GW$PT) that the $GW$ self-energy renormalizes the $e$-ph coupling (from DFPT) in a wide range of materials quantitatively, sometimes even qualitatively \cite{lazzeri2008impact,gruneis2009phonon,faber2011electron,yin2013correlation,antonius2014many,monserrat2016correlation, faber2015exploring, li2021unmasking, li2024two,you2025diverse,alvertis2024influence}, highlighting the fundamental importance of electron correlations in $e$-ph coupling properties. 
To date, most $GW$PT applications focused on metallic systems (e.g., oxide and kagome superconductors) \cite{li2019electron,li2021unmasking,li2024two,you2025diverse,you2025unlikelihood}. 
To extend the applicability of $GW$PT to broad semiconductors and insulators, in particular in conjunction with Wannier interpolation, the question of whether the DFPT-level electrostatic treatment of the Fr\"ohlich interaction suffices for the LR $e$-ph matrix elements from $GW$PT \textit{must} be addressed.
If the semiclassical treatment is insufficient, it becomes imperative to understand and establish how the $GW$ self-energy renormalizes the Fr\"ohlich interaction, and to develop the proper interpolation methodology to incorporate such many-body correlation effects.

In this work, \textit{for the first time}, we unravel the prominent and fundamental many-body correlation effects in polar $e$-ph coupling, and demonstrate the widely adopted semiclassical electrostatic treatment of the Fr\"ohlich interaction is \textit{insufficient}.
Our calculations show that the LO-mode $e$-ph matrix elements from $GW$PT diverge at a different rate than the DFPT matrix elements.
The fundamental contrast resides in that the LO-phonon-induced changes in the $GW$ self-energy diverge at LR ($\textbf{q} \rightarrow 0$), whereas the changes in the exchange-correlation potential in DFT with semi-local functionals do not diverge due to its fictitious SR nature \cite{payne1992iterative}.
We further derive and implement a first-principles-based expression for the $GW$PT LR $e$-ph matrix elements, correctly capturing the asymptotic $GW$PT divergence and seamlessly corroborating with the Wannier interpolation strategy. 
We showcase the critical LR self-energy effects in the phonon-induced electron linewidth, as demonstrated in SiC, BN, GaN, and SrTiO$_3$. Additionally, our calculations reveal considerable $GW$ self-energy effects in the polaron formation of LiF and TiO$_2$ and their nontrivial LR-renormalization behavior.
Remarkably, we demonstrate the necessity of incorporating the full $GW$-level $e$-ph calculation, including the $GW$PT LR contribution, in explaining the photoemission kink in lightly electron-doped TiO$_2$.

\begin{figure}[!t]
\includegraphics[width=1.0\columnwidth]{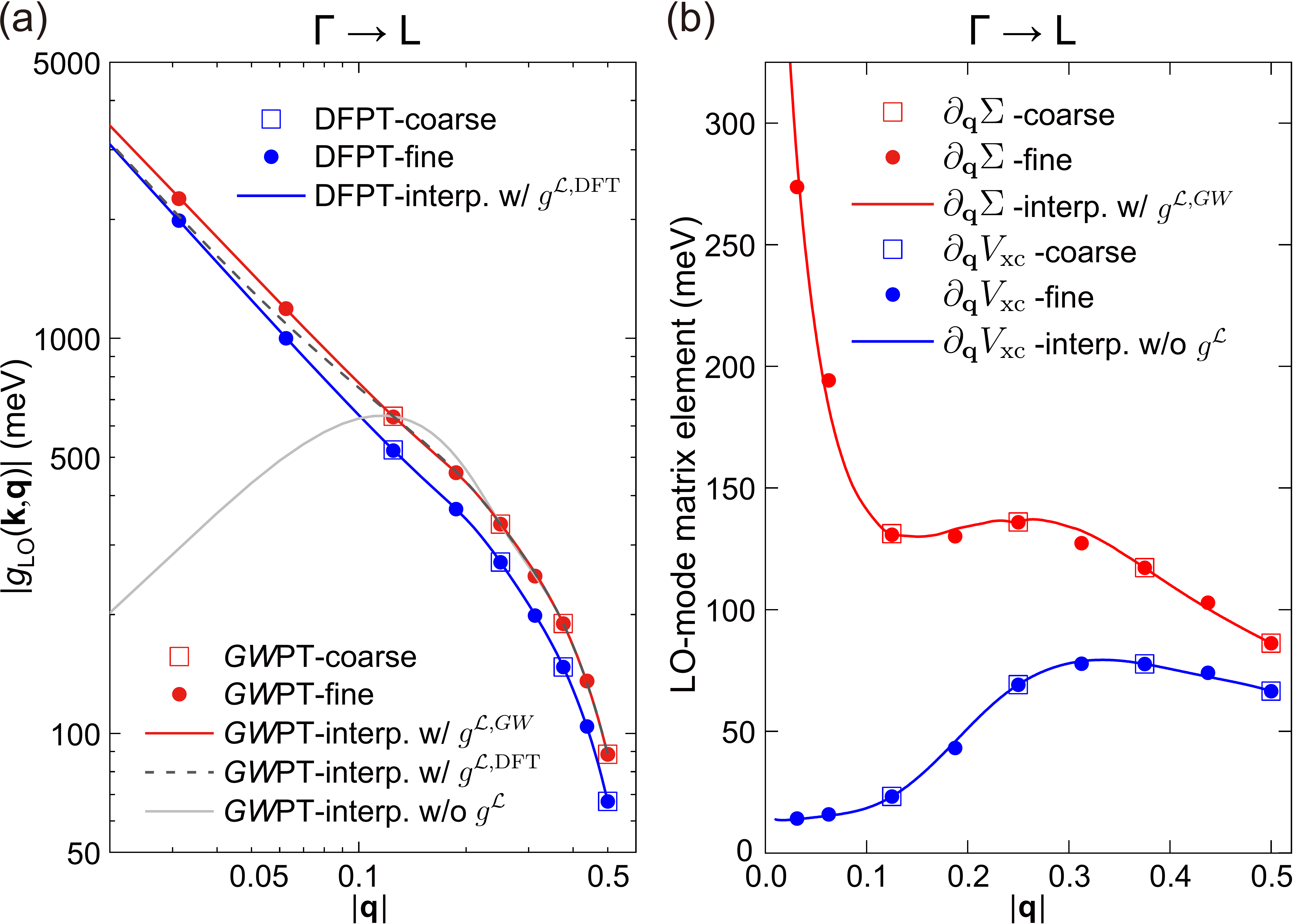}
\caption{\label{fig:1} 
(a) Calculated LO-mode $e$-ph matrix elements (symbols) and the Wannier interpolations (lines) at DFPT (blue color) and $GW$PT (red color) levels in SiC.
The Wannier interpolations are constructed with direct calculations on coarse wavevectors (open squares), and compared against additional direct calculations on fine wavevectors (solid dots).
The gray solid line represents the $GW$PT interpolation without any LR contributions, whereas the gray dashed line represents the $GW$PT interpolation with the DFT-level LR contribution $g^{\mathcal{L}, \text{DFT}}$.
The red solid line shows that by correctly incorporating the $GW$-level LR contribution $g^{\mathcal{L}, GW}$ (Eq.~\eqref{eq:gw-correction}), the interpolation excellently agrees with the direct $GW$PT calculations (red dots) both at the SR regime and towards the LR limit.
(b) Calculated LO-mode matrix elements (symbols) and the Wannier interpolations (lines) for phonon-induced changes in the exchange-correlation potential $\partial_\textbf{q} V_\text{xc}$ (blue color) and in the $GW$ self-energy $\partial_\textbf{q} \Sigma$ (red color).}
\vspace{-10pt}
\end{figure}

We begin our discussion by calculating the $e$-ph matrix elements of the LO phonon mode in the polar semiconductor SiC using DFPT and $GW$PT, which naturally contain both SR and LR contributions directly from first principles. 
Fig.~\ref{fig:1}(a)  shows the asymptotic divergence in the $e$-ph coupling from both DFPT and $GW$PT, however, at different rates. 
Since the computational cost for ultra-fine sampling of the wavevectors of electrons (\textbf{k}-points) and phonons (\textbf{q}-points) is formidable, it has become a well-established standard protocol \cite{verdi2015frohlich, sjakste2015wannier,deng2021ab, sohier2016two} to utilize maximally localized Wannier functions (MLWFs) to interpolate the $e$-ph matrix elements directly calculated on coarse grids. 
The localized nature of MLWFs only allows for interpolating SR interactions, whereas the LR Fr\"ohlich $e$-ph matrix elements are constructed with a dipole-induced electrostatic potential, parametrized by \textit{ab initio} dielectric tensor and Born effective charges from DFPT.
In practice, the LR contribution is first analytically constructed across the Brillouin zone and subtracted from each $e$-ph matrix element on the coarse grid, leaving the remaining part as SR; after the SR part is Wannier-interpolated to a fine grid (or arbitrary wavevectors), the corresponding LR part is analytically calculated and added back \cite{verdi2015frohlich, sjakste2015wannier}.
In Fig.~\ref{fig:1}(a), the well-established interpolation scheme of the DFPT $e$-ph matrix elements excellently agrees with the directly calculated DFPT values, demonstrating that the semiclassical electrostatic treatment of the Fr\"ohlich interaction captures the LR behavior within DFPT.

However, the same electrostatic treatment of the LR Fr\"ohlich interaction is inadequate to capture the LR divergence of the $GW$PT $e$-ph matrix elements. 
As shown in Fig.~\ref{fig:1}(a) with the gray dashed line, while MLWFs can accurately interpolate the SR part (at larger $|\textbf{q}|$) of the $GW$PT $e$-ph matrix elements (similar to the previous applications in metallic systems \cite{li2019electron,li2021unmasking,li2024two,you2025diverse,you2025unlikelihood}), the DFPT-level LR treatment leads to an incorrect divergence tendency as $\textbf{q} \rightarrow 0$  (underestimation by $\sim$ \qty{15}{\percent}), approaching the DFPT limit instead of the $GW$PT one. 
A scrutiny of the discrepancy points to the important many-body correlation effects described by the $GW$ self-energy $\Sigma = iGW$ ($G$: single-particle Green's function; $W$: screened Coulomb interaction) \cite{hedin1970effects, hybertsen1986electron, onida2002electronic}.
As shown in Fig.~\ref{fig:1}(b), at the $GW$PT level, the LO-phonon-induced changes in the electron self-energy $\partial_\textbf{q} \Sigma$ diverge at LR. 
The $GW$ self-energy naturally incorporates the non-local screened Coulomb interaction, thus presenting the $\sim 1/|\textbf{q}|$ singular behavior in the long-wavelength limit.
In strong contrast, the changes to the DFPT-level mean-field exchange-correlation potential $\partial_\textbf{q} V_\text{xc}$ do not diverge within semi-local functionals (e.g., local-density approximation \cite{perdew1981self} and, as in this work, generalized gradient approximation \cite{perdew1996generalized}), missing the true LR divergence. 
The standard semi-local functionals depend only on the local quantities such as the charge density and its gradient, lacking the non-local nature and hence the LR divergent characteristics of the Fr\"ohlich interaction.
In fact, the reason that the semiclassical treatment of the Fr\"ohlich interaction captures the DFPT LR $e$-ph matrix elements is due to the lack of contribution from $\partial_\textbf{q} V_\text{xc}$ as $\textbf{q} \rightarrow 0$, whereas the electrostatic potential reproduces the main effects from the ionic and Hartree potentials \cite{giustino2017electron}.

To reconcile the interpolation of Fr\"ohlich $e$-ph matrix elements at the $GW$PT level, we derive an explicit expression of the self-energy contribution in the LR part. 
Note that $\lim_{\textbf{q}\rightarrow 0} \partial_\textbf{q} V_\text{xc}$ tends to converge to a finite value (Fig.~\ref{fig:1}(b)), and thus it can be treated as a SR contribution and naturally processed via Wannier interpolation. 
Consequently, the $GW$-level LR-part of the $e$-ph matrix element $g^{\mathcal{L}, GW}$ consists of two parts,
\begin{equation} \label{eq:gw-correction}
    g^{\mathcal{L}, GW}_{mn\nu} (\textbf{k}, \textbf{q}; E) = g^{\mathcal{L}, \text{DFT}}_{mn\nu} (\textbf{k}, \textbf{q}) + g^{\mathcal{L}, \Sigma}_{mn\nu} (\textbf{k}, \textbf{q}; E) ,
\end{equation}
where $g^{\mathcal{L}, \text{DFT}}$ denotes the electrostatic contribution at the DFT level as previously developed \cite{verdi2015frohlich, sjakste2015wannier}, $g^{\mathcal{L}, \Sigma}$ denotes the $GW$ self-energy contribution, $m$ and $n$ represent the electron band indices, and $\nu$ labels the phonon branches.
The electron self-energy $\Sigma(E)$ is intrinsically energy-dependent, so are $g^{\mathcal{L},\Sigma}$ and $g^{\mathcal{L}, GW}$.
The $GW$PT $e$-ph matrix elements are constructed within the linear-response theory framework \cite{li2019electron} along with the constant screening approximation \cite{faber2015exploring,li2019electron}. With the single-shot $G_0W_0$PT correction \cite{li2019electron}, we can derive the following effective LR self-energy contribution to $e$-ph matrix elements (see Supplemental Material for derivation details),
\begin{widetext}
\vspace{-6pt}
\begin{equation} \label{eq:long-range-sigma}
\begin{split}
    g^{\mathcal{L}, \Sigma}_{mn\nu} (\textbf{k}, \textbf{q}; E) = 
    & \frac{e^2}{4\pi \epsilon_0 V} \sum_{m'n'} \sum_{\textbf{p}\textbf{G}\textbf{G}'} \frac{4\pi}{|\textbf{p}+\textbf{G}'|^2} \  g^{\mathcal{L}, \text{DFT}}_{m'n'\nu} (\textbf{k}-\textbf{p}, \textbf{q}) 
    \braket{\psi_{m\textbf{k}+\textbf{q}} | e^{i(\textbf{p}+\textbf{G})\cdot \textbf{r}} | \psi_{m'\textbf{k}+\textbf{q}-\textbf{p}} }
    \braket{\psi_{n'\textbf{k}-\textbf{p}} | e^{-i(\textbf{p}+\textbf{G}')\cdot \textbf{r}'} | \psi_{n\textbf{k}} }   \\[2ex]
    & \times \frac{F(\varepsilon_{n'\textbf{k}-\textbf{p}}; E) - F(\varepsilon_{m'\textbf{k}+\textbf{q}-\textbf{p}}; E) }{\varepsilon_{n'\textbf{k}-\textbf{p}} - \varepsilon_{m'\textbf{k}+\textbf{q}-\textbf{p}}}  ,
\end{split}
\raisetag{20pt}
\end{equation}
\vspace{-15pt}
\end{widetext}
where the function $F$ is defined as,
\begin{equation} \label{eq:F function}
    F(\varepsilon_{n\textbf{k}}; E) =
    \begin{cases}
        \dfrac{\omega_p}{2\alpha (E - \varepsilon_{n\textbf{k}} + \alpha \omega_p)} - 1, \text{ if } n \leqslant N_v,\\[2ex]
        \dfrac{\omega_p}{2\alpha (E - \varepsilon_{n\textbf{k}} - \alpha \omega_p)}, \text{\ \ \ \ \ \  if } n > N_v,
    \end{cases}
\end{equation}
with,
\begin{equation} \label{eq:alpha}
    \alpha = \left[  1 - \epsilon_{00}^{-1}(\textbf{p}=0, \omega=0)  \right]^{-\frac{1}{2}} .
\end{equation}
In Eqs. \eqref{eq:long-range-sigma}-\eqref{eq:alpha}, $\epsilon_0$ is the vacuum permittivity, $V$ is the volume of the Born-von K\'arm\'an (BvK) supercell, $\varepsilon_{n\textbf{k}}$ is the electron band energy, $\omega_p$ is the plasmon frequency, $N_v$ is the number of valence bands, and $\epsilon_{00}^{-1}$ represents the head element ($\textbf{G}=0$, $\textbf{G}'=0$) of the inverse dielectric matrix $\epsilon^{-1}_{\textbf{G}\textbf{G}'}(\textbf{p}, \omega)$. 
In arriving at Eqs. \eqref{eq:long-range-sigma}-\eqref{eq:alpha}, we treat the frequency dependence of the inverse dielectric matrix via the generalized plasmon-pole model \cite{hybertsen1985first, hybertsen1986electron,zhang1989evaluation,deslippe2012berkeleygw} and further neglect the local-field effects as we focus on the LR $\textbf{q}\rightarrow 0$ regime, where details within the primitive unit cell become less relevant (see Supplemental Material for more discussions). 
In Eq.~\eqref{eq:long-range-sigma}, we introduce a small non-zero imaginary part to the denominator to avoid the numerical instability from the degenerate subspace in the summation over states \cite{li2019electron, ponce2015temperature}.
We notice that the energy dependence of $\Sigma(E)$ is relatively weak and therefore we take an average energy $E = \frac{1}{2}(\varepsilon_{m\textbf{k}+\textbf{q}} + \varepsilon_{n\textbf{k}})$ for evaluating the matrix element $g_{mn\nu}(\textbf{k}, \textbf{q})$. 
%The mean-field DFT and DFPT calculations are carried out with norm-conserving pseudopotentials in the generalized gradient approximation \cite{perdew1996generalized}. 
This formalism can be straightforwardly generalized to include quadrupolar $e$-ph coupling terms \cite{jhalani2020piezoelectric, park2020long, brunin2020electron, brunin2020phonon}.

We implement Eqs. \eqref{eq:gw-correction}-\eqref{eq:alpha} into the Wannier interpolation workflow of the $e$-ph coupling \cite{li2024electron, giustino2007electron, ponce2016epw, lee2023electron} to capture the many-body LR correlation effects with low computational overhead. 
The workflow can be represented in a  simplified notation: DFT $\rightarrow$ DFPT $\rightarrow GW \rightarrow GW\text{PT} \rightarrow$ Wannier SR interpolation with LR treatment.
As shown in Fig.~\ref{fig:1}, combined with the $GW$PT LR contribution, the interpolated $e$-ph matrix elements (red lines) are in excellent agreement with the directly calculated $GW$PT values (red solid dots), highlighting the accuracy of our approach.
In the Supplemental Material, we further validate this new method on cubic BN, wurtzite GaN, and cubic SrTiO$_3$, demonstrating its general applicability and robustness.

\begin{figure}[!t]
\includegraphics[width=1.0\columnwidth]{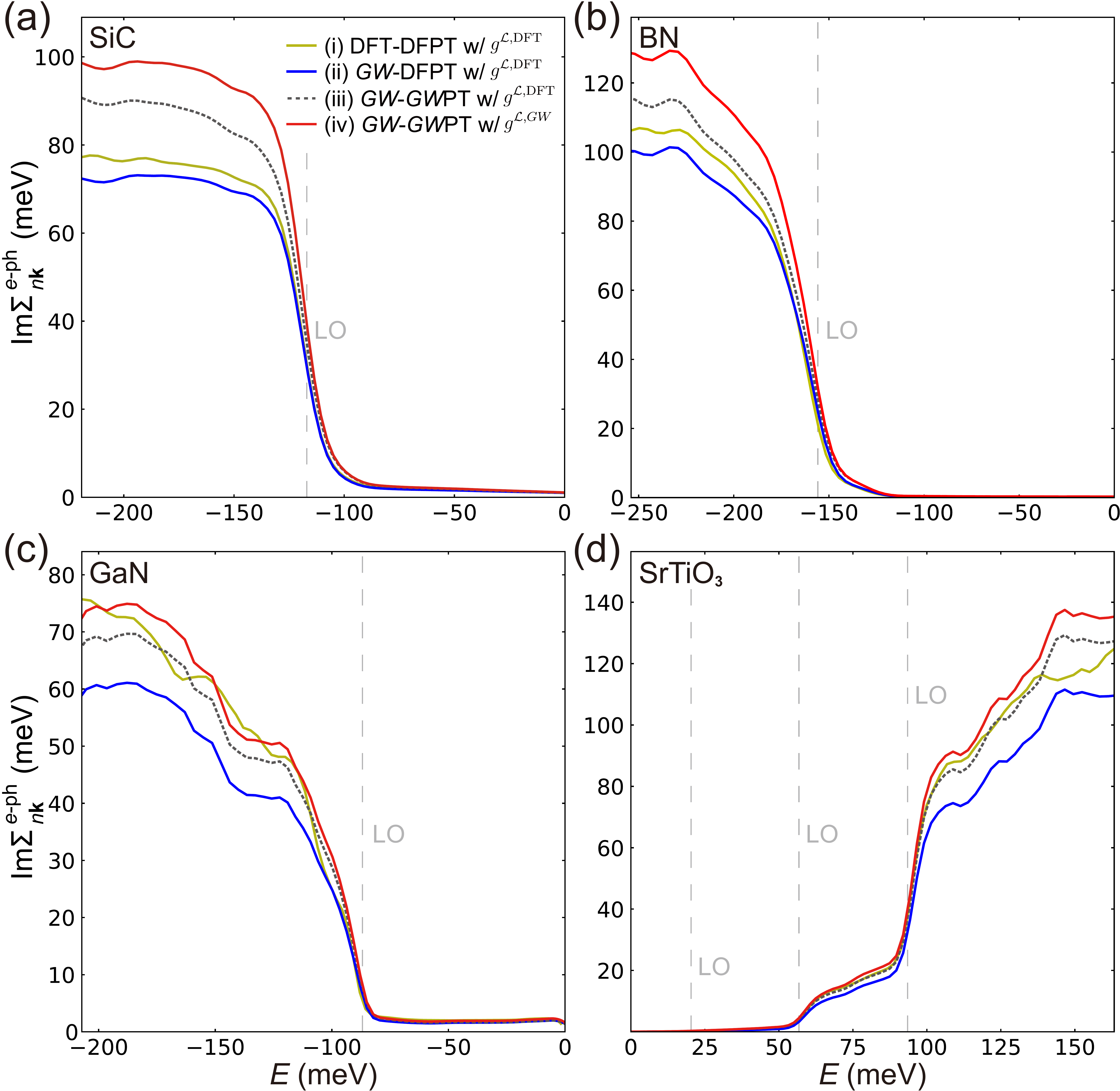}
\caption{\label{fig:2}
Calculated imaginary part of the phonon-induced electron self-energy $\text{Im}\Sigma^{e\text{-ph}}_{n\textbf{k}}$ at 20 K. (a)-(c) Hole linewidth near the VBM for SiC, cubic BN, and wurtzite GaN, respectively. (d) Electron linewidth near the CBM for SrTiO$_3$.
Vertical gray lines indicate the characteristic LO phonon energies.
Four levels of theory are compared: (i) the standard DFT and DFPT baseline, \rev{with the $e$-ph matrix elements containing the conventional DFT-level SR and LR parts}; (ii) $GW$ quasiparticle energies combined with conventional DFPT $e$-ph matrix; (iii) $GW$ eigenvalues with $GW$PT corrections only in the SR $e$-ph matrix elements, whereas the LR contribution is accounted for at the DFPT level ($g^{\mathcal{L},\text{DFT}}$); and (iv) the fully $GW$ and $GW$PT treatment, where the LR $e$-ph elements are treated also at the $GW$ level ($g^{\mathcal{L},GW}$). 
 }
\vspace{-10pt}
\end{figure}

To elucidate the many-body correlation effects on physical observables, we first examine the electron linewidth through the imaginary part of the Fan-Migdal phonon-induced electron self-energy Im$\Sigma^{e\text{-ph}}_{n\textbf{k}}$ \cite{migdal1958interaction}. We investigate hole states near the valence band maximum (VBM) of SiC (Fig.~\ref{fig:2}(a)), BN (Fig.~\ref{fig:2}(b)), and GaN (Fig.~\ref{fig:2}(c)), as well as electron states near the conduction band minimum (CBM) of SrTiO$_3$ (Fig.~\ref{fig:2}(d)). 
In the limit of infinitesimal doping concentrations, where the Fermi level $E_F$ locates at the band edges, the linewidth of the low-energy excitations is contributed by small-$|\textbf{q}|$ phonon scattering. The LO phonon modes with divergent $e$-ph matrix elements thus dominate the linewidth. 
Consequently, for excitation energies below the LO phonon energy, i.e., $|E-E_F| < \hbar\omega_{\text{LO}}$, the linewidth (or Im$\Sigma^{e\text{-ph}}_{n\textbf{k}}$) is low, whereas when near the LO phonon frequencies, the linewidth rapidly rises, as shown in Fig.~\ref{fig:2}.
The linewidth onset near the LO phonon frequencies is thus very sensitive to the LR part of the $e$-ph coupling.

To isolate the effects of $GW$ quasiparticle band renormalization and the $GW$PT corrections to the $e$-ph matrix elements including both the SR and LR contributions, we compare four levels of theory: (i) the standard DFT and DFPT baseline ($\varepsilon^\text{DFT}$, \rev{$g^{\mathcal{S},\text{DFT}}+ g^{\mathcal{L},\text{DFT}}$}) \rev{with the conventional DFPT LR limit $g^{\mathcal{L},\text{DFT}}$ (the first term of Eq. \eqref{eq:gw-correction})}; (ii) $GW$ quasiparticle energies combined with \rev{the same} conventional DFPT $e$-ph matrix elements ($\varepsilon^{GW}$, \rev{ $g^{\mathcal{S},\text{DFT}}+ g^{\mathcal{L},\text{DFT}}$ }); (iii) $GW$ eigenvalues with $GW$PT $e$-ph matrix elements but retaining the DFPT LR limit ($\varepsilon^{GW}$, $g^{\mathcal{S},GW} + g^{\mathcal{L},\text{DFT}}$); and (iv) the fully $GW$-level treatment ($\varepsilon^{GW}$, $g^{\mathcal{S},GW} + g^{\mathcal{L},GW}$).
As shown in Fig.~\ref{fig:2}, while the $GW$ renormalization in the quasiparticle band structures tends to reduce the linewidth ((i) \textit{vs.} (ii)) by suppressing density of states near band edges, $GW$PT further enhances the $e$-ph matrix elements compared with DFPT, leading to larger linewidths ((ii) \textit{vs.} (iv)).
Particularly for the LR $e$-ph coupling, the clear distinction between the standard electrostatic treatment of $g^{\mathcal{L}, \text{DFT}}$ and our many-body expression of $g^{\mathcal{L}, GW}$ ((iii) \textit{vs.} (iv)) underscores the importance of correlation effects in the Fr\"ohlich interaction, which is absent in the traditional semiclassical treatment.

\begin{figure}[!thb]
\includegraphics[width=1.0\columnwidth]{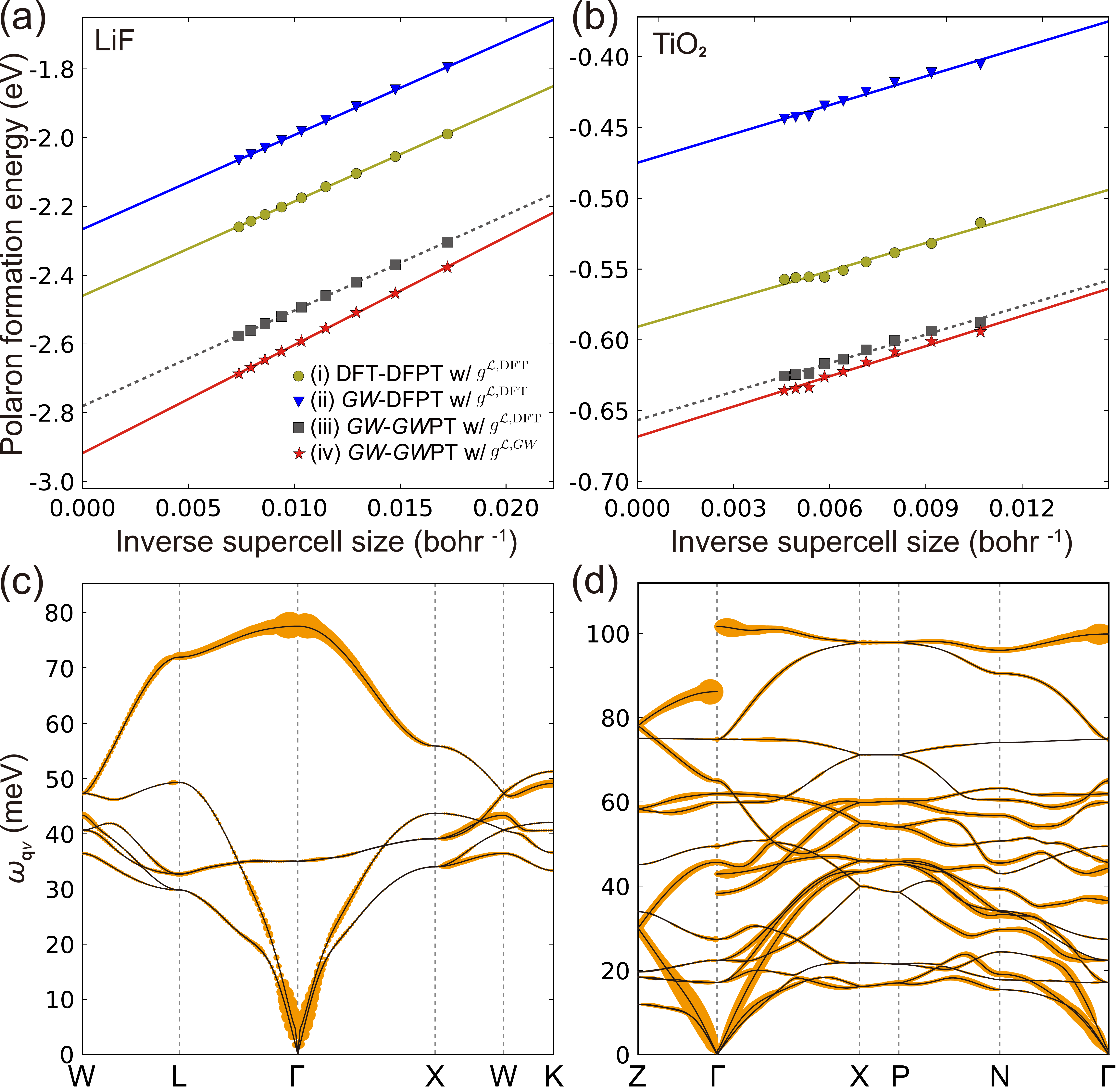}
\caption{\label{fig:3} Hole polaron formation energy $\Delta E_f$ (symbols) as a function of inverse BvK supercell size and the extrapolation (dotted lines) to the dilute limit (infinite supercell size) for (a) LiF and (b) TiO$_2$. 
Four levels of theory are presented.
Phonon dispersion of (c) LiF and (d) TiO$_2$ overlaid by spectral decomposition of the lattice distortion at the full $GW$ and $GW$PT level (case (iv)).
}
\vspace{-10pt}
\end{figure}

In the following, we further investigate the $GW$ and $GW$PT self-energy effects in the polaron formation processes, particularly from the LR part $g^{\mathcal{L},GW}$.
A polaron is a quasiparticle formed with an electron or a hole dressed by a cloud of lattice distortions \cite{alexandrov2008polarons, devreese2009frohlich, alexandrov2010advances, franchini2021polarons}, fundamentally governed by the $e$-ph interaction and having profound influence on the transport and optical properties \cite{caruso2021two,zhang2023bipolaronic,wang2023giant,celiberti2024spin,pavlyukh2022time, reticcioli2019interplay,smart2019optical}. 
Besides standard supercell approaches based on DFT \cite{spreafico2014nature, kokott2018first}, new $\textit{ab initio}$ methods of polaron equations have been recently developed to compute the polaron formation energy $\Delta E_f$ and wavefunctions, using linear-response $e$-ph ingredients along with Wannier interpolation~\cite{sio2019polarons, sio2019ab, vasilchenko2022variational, vasilchenko2025variational}.
%{\color{red} Other significant progress in the theory of polarons include renormalization group \cite{grusdt2016all},  canonical transformations \cite{lee2021facile, luo2022comparison}, Green’s function methods \cite{lafuente2022unified, lafuente2022ab, houtput2025first}, path-integral Monte Carlo\cite{ houtput2021beyond} and diagrammatic Monte Carlo methods \cite{mishchenko2000diagrammatic, prokof2008fermi, luo2025first}.}
In addition, first-principles diagrammatic Monte Carlo methods have been developed to treat the many-body polaron problem to all orders~\cite{luo2025first}.
The existing first-principles studies of polarons, however, predominantly rely on DFT eigenvalues and DFPT $e$-ph matrix elements, missing the important many-electron self-energy effects \cite{freysoldt2014first}.
%{\color{red} and also suffer from the self-interaction error \cite{perdew1981self, mori2008localization, sadigh2015variational, falletta2022polarons, dai2025comparison} coming with the exchange-correlation potential}
Here, we go beyond the previous DFT-level polaron studies by performing full $GW$-level calculations, along with incorporating the many-body $GW$PT LR Fr\"ohlich interaction.

Fig.~\ref{fig:3}(a) and (b) plot the hole polaron formation energy $\Delta E_f$ of LiF and anatase TiO$_2$, as a function of supercell size (defined by the \textbf{k}- and \textbf{q}-point mesh size) by solving the polaron equations ~\cite{sio2019polarons, sio2019ab}. The polaron equations built with perturbative $e$-ph matrix elements have been demonstrated to well capture the main characteristics of polarons across a wide range of materials \cite{dai2026comparison}.
With a similar setup of the four levels of theory (as presented and discussed for linewidth (Fig.~\ref{fig:2})), we notice that the $GW$ renormalization in the band structure reduces the magnitude of the formation energy $|\Delta E_f|$ ((i) \textit{vs.} (ii)), whereas $GW$PT significantly enhances the $e$-ph coupling and leads to a larger $|\Delta E_f|$ ((ii) \textit{vs.} (iv)). Importantly, we note that while the enhancement factor in the $e$-ph matrix elements themselves $|g^{GW}|/|g^\text{DFT}|$ is moderate $\lesssim 1.1$ for the relevant states in LiF and TiO$_2$, the enhancement factor in $|\Delta E_f|$ is significantly larger $\sim 1.25 - 1.4$ (from case (ii) to case (iv)), highlighting the nonlinear behaviors in the solutions of the self-consistent coupled polaron equations ~\cite{sio2019polarons, sio2019ab}.

\begin{figure}[!thb]
\includegraphics[width=1.0\columnwidth]{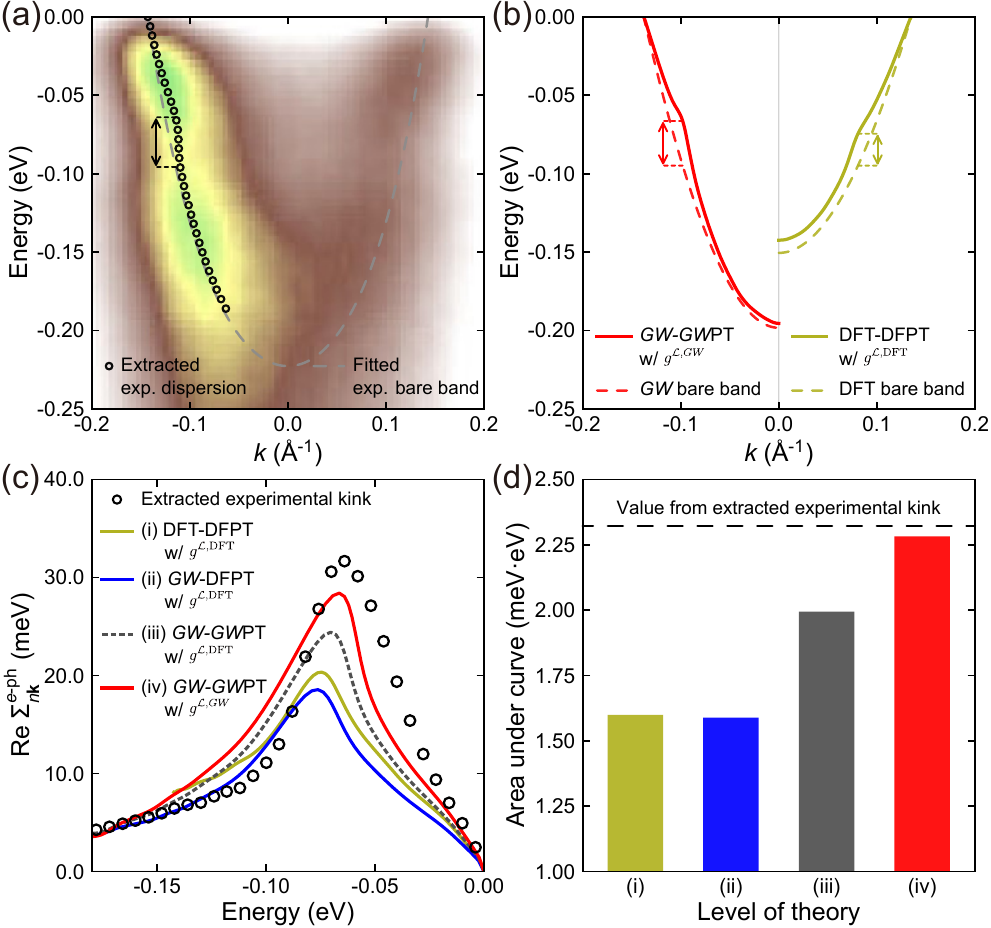}
\caption{\label{fig:4} (a) The experimental ARPES spectrum from Ref. \cite{moser2013tunable} with electron carrier concentration $3.5\times{10}^{20}$ cm$^{-3}$ at 20 K, overlaid by the extracted dispersion (black circles) and the fitted bare band (gray dashed line) (see Supplemental Material \rev{for extraction and fitting details}). (b) The renormalized quasiparticle dispersion relations from the computed spectral function and bare bands, from full $GW$-level (left, red) and full DFT-level (right, yellow) calculations, at the same doping level and temperature. Double-headed arrows in (a) and (b) indicate the kinks. (c)  $\text{Re}\Sigma_{n\textbf{k}}^\text{$e$-ph}$ as a function of the quasiparticle excitation energy $E_{n\textbf{k}}$. \rev{The circles represent the extracted  experimental kink based on the extracted dispersion relation and the fitted bare band in (a).} (d) The kink strength indicated by the area under curve from (c) with an integration range from 0 meV to $-180$ meV (case (i) data set ends at $-142.4$ meV).  \rev{The value from extracted experimental kink} is represented by the horizontal dashed line. 
}
\vspace{-10pt}
\end{figure}

Next, we focus on the impact from the LR $e$-ph coupling $g^{\mathcal{L}, GW}$ compared with $g^{\mathcal{L},\text{DFT}}$ in the polaron formation. In LiF hole polaron (Fig.~\ref{fig:3}(a)), $g^{\mathcal{L},GW}$ considerably alters the position and the slope of the extrapolation line from $g^{\mathcal{L},\text{DFT}}$ ((iii) \textit{vs.} (iv)). 
In contrast, this renormalization is less pronounced for TiO$_2$ hole polaron (Fig.~\ref{fig:3}(b)). 
The different behavior in the two material systems can be traced back to the spectral decomposition of the lattice distortion, as shown in Fig.~\ref{fig:3}(c) and (d) (with the circle area proportional to $|B_{\textbf{q}\nu}|^2$ defined in Refs.~\cite{sio2019polarons, sio2019ab}), which represent the degree of involvement of different phonon modes in forming the polaron. 
We note that the LiF hole polaron is predominantly driven by the LO mode near the $\Gamma$ point, where the $\sim 1/|\mathbf{q}|$ singularity and the corresponding $GW$PT LR correction are crucial. On the other hand, the TiO$_2$ hole polaron involves a broader range of phonon modes (across wavevectors and branches), effectively diluting the role of the LR contribution.

Last but not least, we demonstrate the impact of the $GW$ self-energy renormalization in $e$-ph coupling, particularly the LR part, reflected in the spectroscopic features of lightly doped semiconductors. Angle-resolved photoemission spectroscopy (ARPES) has revealed a strong photoemission kink in electron-doped anatase TiO$_2$ \cite{moser2013tunable}. 
Previous first-principles studies based on DFT and DFPT focused on the signature of polarons at low carrier densities \cite{verdi2017origin, caruso2018electron}, but predicted a much weaker photoemission kink in TiO$_2$, leaving this important spectroscopic feature unresolved.

We construct the spectral function of doped TiO$_2$ by computing the Fan-Migdal $e$-ph self-energy along with the cumulant expansion \cite{verdi2017origin}. 
The screening of the $e$-ph matrix elements arising from the doped carriers is accounted for \cite{verdi2017origin}. 
In Fig.~\ref{fig:4}(a) and (b), we compare directly the dispersion relation \rev{extracted from the experiment \cite{moser2013tunable} with the theoretical calculation}, for the electron carrier concentration of $3.5\times10^{20}$  cm$^{-3}$ (see Supplemental Material for \rev{details of the experimental data extraction procedure}).
We notice that the $GW$ renormalization leads to higher Fermi velocity and a lower CBM, showing an improved agreement with experiment compared with DFT results. Then, we compare the real part of the phonon-induced electron self-energy
$\text{Re}\Sigma_{n\textbf{k}}^\text{$e$-ph}$,
which directly induces the dispersion kink at $\sim -70$ meV.
Fig.~\ref{fig:4}(c) shows the direct comparison of $\text{Re}\Sigma_{n\textbf{k}}^\text{$e$-ph}$ from the four levels of theory, along with \rev{the one extracted from} the experiment. 
We find that the $GW$PT matrix elements substantially enhance the predicted kink magnitude.
Particularly, the $GW$PT LR Fr\"ohlich contribution is crucial in obtaining a better agreement in peak height with experiment ((iii) \textit{vs.} (iv)). 
Moreover, we perform an area under curve integration for the data in Fig.~\ref{fig:4}(c) as an indicator for the overall $e$-ph coupling strength (mitigating experimental uncertainty and theoretical errors in phonon frequencies and approximations adopted). 
As shown in Fig.~\ref{fig:4}(d), case (iv) agrees with the experiment kink area nicely, where the many-body correlation effect contributes significantly to the overall $e$-ph coupling strength.

In summary, this work reveals the previously unrecognized and important many-electron self-energy effects in the LR Fr\"ohlich-type $e$-ph coupling.
We develop and implement a many-body approach in seamless integration with the Wannier interpolation technique for describing the Fr\"ohlich LR $e$-ph coupling with the $GW$ self-energy effects incorporated. \rev{These many-body effects can lead to substantial corrections beyond the conventional DFT-level description and have a broad and strong relevance to essentially all compound semiconductors and insulators.}
Our work enables a comprehensive first-principles workflow to study the $e$-ph coupling phenomena in compound semiconductors and insulators at the full $GW$ and $GW$PT level with an unprecedented many-body accuracy.

\textit{Acknowledgement.} This work was primarily supported by the U.S. National Science Foundation (NSF) CAREER Award under Grant No. DMR-2440763, for the development of the theory and computational method, and for the calculations. 
The integration and interfacing of the method into community codes \textsc{BerkeleyGW} and \textsc{EPW} were supported by the U.S. NSF under Grant No. OAC-2513830.
C.E.H. and H.C.H. acknowledge the funding support of the National Science and Technology Council, Taiwan, under Grant Numbers NSTC 114‑2112‑M‑032‑009‑MY3 and NSTC 114-2811-M-032-012-MY2.
M.D.B. acknowledges the support by the Center for Computational Study of Excited-State Phenomena in Energy Materials (C2SEPEM) at the Lawrence Berkeley National Laboratory (LBNL), which is funded by the U.S. Department of Energy (DOE), Office of Science (SC), Basic Energy Sciences, Materials Sciences and Engineering Division under Contract No. DE-AC02-05CH11231, as part of the Computational Materials Sciences Program.
A.M.A. thanks the Physics Department and the Oden Institute for Computational Engineering and Sciences at The University of Texas at Austin for their support.
An award of computer time was provided by the U.S. DOE INCITE program. This research used computational resources of both the Argonne and Oak Ridge Leadership Computing Facilities, which are U.S. DOE SC User Facilities supported under contracts DE-AC02-06CH11357 and DE-AC05-00OR22725.
Computational resources were also provided by National Energy Research Scientific Computing Center, which is a U.S. DOE SC User Facility supported under contract DE-AC02-05CH11231, and by Texas Advanced Computing Center, which is supported by U.S. NSF under Grant No. OAC-1818253.

\bibliography{ref}

\end{document}